\documentclass{PoS}

\title{Hadron interactions from lattice QCD}

\ShortTitle{Hadron interactions from lattice QCD}

\author{%\speaker{
Sinya Aoki\\%}\thanks{A footnote may follow.}\\
        Graduate School of Pure and Applied Sciences, University of Tsukuba, Tsukuba, Ibaraki 305-8571, Japan\\
        Riken BNL Research Center, Brookhaven National Laboratory, Upton, NY 11973, USA\\
        E-mail: \email{saoki@het.ph.tsukuba.ac.jp}}

\abstract{Studies  on hadron interactions from lattice QCD are reviewed. 
The $S$-wave $\pi\pi$ scattering lengths of the $I=0$ and $I=2$ channels are extracted from
various lattice determinations of low energy constants in $N_f=2$ chiral perturbation theory.
The results agree with each other and agree also with other non-lattice estimates.
Recently
the $P$-wave $\pi\pi$ scattering phase shift for the $I=1$ channel has been calculated.
A preliminary estimate of the $\rho$ meson decay width from the phase shift is consistent with the experimental value. 
Two approaches to potentials between hadrons are discussed. One is a method using static quarks to define the distance between two hadrons. The other is a method to define a potential from a wave function of two hadrons.  An application of the latter to the nucleon-nucleon ($NN$) potential turns out to reproduce qualitative features of the phenomenological $NN$ potential such as attraction at long distance
and repulsion at short distance. Theoretical issues of this approach are also discussed. }

\FullConference{The XXV International Symposium on Lattice Field Theory\\
		 July 30-4 August 2007\\
		 Regensburg, Germany}

\begin{document}

\section{Introduction}
As numerical simulations in lattice QCD mature, we can attack more difficult or complicated problems in strong interactions. Calculations of hadronic matrix elements of electroweak operators are predictions from lattice QCD, which can be used to determine parameters of the standard model. Another interesting application is to investigate interactions among hadrons using lattice QCD.
Since hadrons are bound states of quarks and gluons, their interactions are residual effects of strong interactions among them inside hadrons, and therefore they are more complicated quantities in lattice QCD than properties of an isolated hadron such as masses or decay constants.
In addition, the euclidean nature of lattice QCD makes problems of hadron interactions more complicated and difficult. Thanks to the finite volume technique\cite{Luescher1}, however, scattering lengths and phase shifts of hadrons can in principle be extracted from lattice QCD. Some recent progress on these quantities is reviewed in this paper.

\begin{figure}[b]
\centering
\includegraphics[width=80mm,clip]{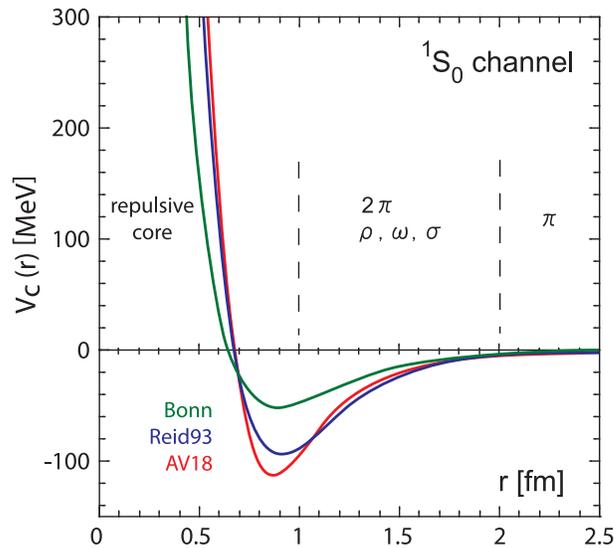}
\hspace*{5mm}
\caption{Three examples of the modern $NN$ potential in $^1S_0$ (spin singlet and $S$-wave) channel: Bonn\cite{Bonn}, Reid93\cite{Reid93} and AV18\cite{AV18}.}
\label{fig:potential}
\end{figure}

I also consider a more complicated but important quantity of hadron interactions, the force between nucleons (the nuclear force). In 1935 Yukawa introduced a virtual particle, the pion, to account for the nuclear force\cite{Yukawa}, by which protons and neutrons are confined in nuclei. Since then  enormous efforts have been made to understand the nucleon-nucleon ($NN$) interaction at low energies both theoretically and experimentally. In Fig.~\ref{fig:potential},  I present modern $NN$ potentials, which are characterized by the following features\cite{Taketani, Machleidt}.  At long distances ($r\ge 2$ fm ) there exists weak attraction, which
is well understood and is dominated by the one pion exchange (OPE), as first pointed out by Yukawa.
At medium distances (1 fm $\le r \le $ 2 fm), contributions from the exchange of multi-pions and heavy mesons such as $\rho$, $\omega$ and $\sigma$ lead to slightly stronger attraction.
At short distances ($r \le$ 1 fm), attraction turns into repulsion, and it becomes stronger and stronger as $r$ gets smaller, forming the strong repulsive core\cite{Jastrow}, which is essential not only for describing the $NN$ scattering data, but also for the stability and saturation of atomic nuclei, for determining the maximum mass of neutron stars, and for igniting Type II supernova explosions\cite{Supernova}.
Although the origin of the repulsive core must be related to the quark-gluon structure of nucleons,
it remains one of the most fundamental problems in nuclear physics for a long time\cite{OSY}.
It is a great challenge for us to derive the nuclear potential including the repulsive core from lattice QCD.
Recent progress on this issue is explained in this paper.

\section{Conventional method in lattice QCD}
In this section, some recent results for the  scattering length and phase shift for pions are discussed. 

\subsection{Scattering length of pions}
\begin{figure}[t]
\centering
\includegraphics[width=80mm,angle=270]{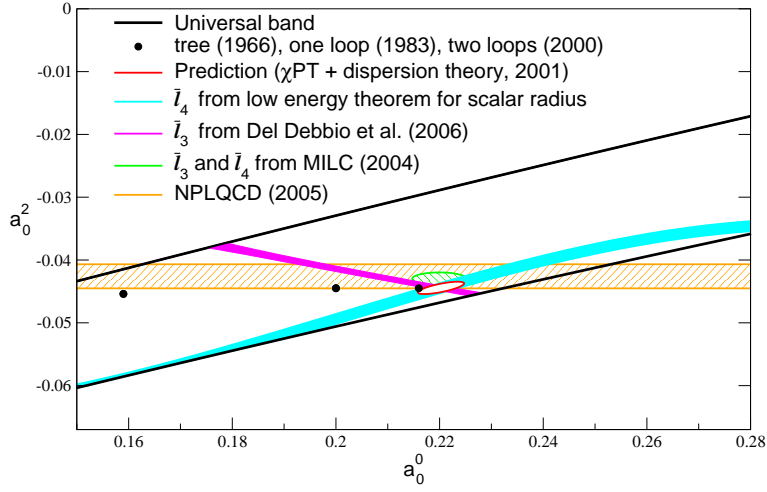}
\caption{$S$-wave scattering lengths in unit of $1/M_\pi$.}
\label{fig:a0a2}
\end{figure}

In Fig.~\ref{fig:a0a2}, recent theoretical estimates of the $S$-wave $\pi\pi$ scattering lengths, $a_0^0$ and $a_0^2$ in units of $1/M_\pi$, are summarized \cite{Leutwyler}, where $L$ and $I$ of the scattering length $a_L^I$ represent the total angular momentum and the total isospin, respectively. The band bounded by two black solid lines represents the theoretically allowed region (universal band)\cite{Leutwyler}. The three black dots are predictions from 2-flavor chiral perturbation theory (ChPT) at tree, 1-loop and 2-loop levels from left to right. Other results are obtained by using the formulae\cite{CGL}
\begin{eqnarray}
M_\pi (2a_0^0 + 7 a_0^2) &=& -6\pi \left(\frac{M_\pi}{4\pi F_\pi}\right)^4 \left(\bar l_3 -\frac{193}{210}\right)+
M_\pi^4\alpha_3 + O(M_\pi^6), \\
M_\pi (2a_0^0 - 5 a_0^2) &=& \frac{3M_\pi^2}{4\pi F_\pi^2}+24\pi \left(\frac{M_\pi}{4\pi F_\pi}\right)^4 \left(\bar l_4 -\frac{887}{840}\right)
+M_\pi^4\alpha_4 + O(M_\pi^6), 
\end{eqnarray}
where $M_\pi^4\alpha_{3,4}$ are corrections to ChPT from dispersion integrals. They are approximated at the physical point as
\begin{eqnarray}
M_\pi^4\alpha_3 &=& 0.135 +0.77(a_0^0 - 0.220) - 1.50(a_0^2 + 0.0444),\\
M_\pi^4\alpha_4 &=& 0.061 +0.48(a_0^0 - 0.220) - 0.26(a_0^2 + 0.0444) .
\end{eqnarray}
$\bar l_{3,4}$ are the low energy constants in $N_f=2$ ChPT and defined by
\begin{eqnarray}
M_\pi^2 &=& M^2 \{ 1+ \frac{x}{2}\bar l_3 + O(x^2)\}, \quad
F_\pi = F\{ 1+ x \bar l_4 + O(x^2)\} 
\end{eqnarray}
with $M^2 = 2 B m$ and $x \equiv M^2/(16\pi^2 F^2)$. Here $m$ is the quark mass and $F$ is the pion decay constant in the chiral limit. Once $\bar l_3$ and $\bar l_4$ are known, the above expressions give $a_0^0$ and $a_0^2$.

 The red ellipse in the figure corresponds to  $\bar l_3 = 2.9(2.4)$ from the mass spectrum of the pseudoscalar octet and $\bar l_4 = 4.4(2)$ from the scalar form factor of the pion.  
The narrow azure strip indicates the region allowed if $\bar l_3$ is treated as free parameter while
$\bar l_4$ is fixed to the above value.
There exist several lattice determinations of $\bar l_3$ and $\bar l_4$. The MILC collaboration\cite{MILC}, 
using 2+1 flavors of staggered dynamical quarks, obtain the low energy constants $L_{4,5,6,8}$ of $N_f=3$
ChPT, which are translated to $\bar l_3= 0.8(2.3)$ and $\bar l_4 = 4.0 (6)$ by standard one loop formulae. The results for $a_0^0$ and $a_0^2$ are  given by a green ellipse in the figure, which is consistent with the previous estimate (the red ellipse).
The result of $N_f=2$ dynamical Wilson quarks, $\bar l_3 = 3.0(5)(1)$\cite{DGLPT}, with free $\bar l_4$  leads to the narrow purple strip in the figure, which intersects the red ellipse.

Recent results of the ETM collaboration\cite{ETMC}, $\bar l_3 = 3.65(12)$ and $\bar l_4 = 4.52(6)$,
give $M_\pi a_0^0 =0.221(1)$ and $M_\pi a_0^2 = - 0.0427(1)$ (with $M_\pi = 140$ MeV and $F_\pi =93$ MeV), which nicely agree with the red ellipse with smaller errors.

These lattice results for $a_0^0$ and $a_0^2$ are indirectly obtained by using $\bar l_3$ and $\bar l_4$, which have been determined from the quark mass dependences of $M_\pi$ and $F_\pi$. There exist several direct estimates on $a_0^2$ using the finite volume method\cite{Luescher1}. 
The horizontal orange band in Fig.~\ref{fig:a0a2} indicates the direct result for $a_0^2$ by
the NPLQCD collaboration\cite{NPLQCD1} with domain-wall valence quarks on a 2+1 staggered sea.
Recently they have reported a more accurate result, $M_\pi a_0^2 = - 0.0443(4)$\cite{NPLQCD2},
which is consistent with the red ellipse. So far ChPT predictions and all indirect /direct lattice results agree with each other. However a direct evaluation of $a_0^0$ is missing. Thus it is a great challenge for lattice QCD to extract $a_0^0$ directly by the finite volume method.
 
 \subsection{The $\rho$ meson decay width}
 A less difficult than $a_0^0$ but still challenging problem is the determination of the $P$-wave scattering phase shift for the $I=1$ two-pion system, from which the $\rho$ meson decay width is extracted. 
 Applying the finite volume method in the laboratory system to this problem,
 the CP-PACS collaboration has calculated the decay width\cite{Ishizuka0} in lattice QCD with  a renormalization group improved gauge action and $N_f=2$ dynamical clover quarks on a $12^3\times 24$ lattice at $m_\pi/m_\rho \simeq0.41$ and the lattice spacing $1/a \simeq 0.92$ GeV. In order to realize the kinematics such that the energy of the two pions is close to $m_\rho$, one pion has a non-zero momentum ${\bf p}=(2\pi/L) {\bf e}_3$ and the other is at rest, while the $\rho$ meson has the same non-zero momentum. Energies of these states for non-interacting hadrons are $W_1^0 = \sqrt{m_\pi^2 + p^2} + m_\pi$ for the two pions and $W_2^0 = \sqrt{m_\rho^2 +p^2}$ for the $\rho$. At $m_\pi/m_\rho \simeq 0.41$ on a $12^3\times 24$ lattice, the invariant mass of two free pions takes the value $\sqrt{s} \simeq 0.97 \times m_\rho$, which is much closer to $m_\rho$ than 
 that in the center of mass system, $E = 2\sqrt{m_\pi^2 + p^2} \simeq1.47 m_\rho$.
 The hadron interaction shifts the energies from $W_n^0$ to $W_n$, which are related to the two-pion scattering phase shift $\delta$ in the infinite volume through the Rummukainen-Gottlieb formula\cite{RG}.
 
 To extract two energy levels $W_n$ close to each other, a $2\times 2$ matrix of the time correlation function
\begin{equation}
G(t) = \left(
\begin{array}{cc}
\langle 0 \vert (\pi\pi)^\dagger (t) (\pi\pi) (t_s)\vert 0 \rangle & \langle 0 \vert (\pi\pi)^\dagger (t) \rho_3 (t_s)\vert 0 \rangle \\
\langle 0 \vert  \rho_3^\dagger (t)(\pi\pi) (t_s) \vert 0 \rangle &  \langle 0 \vert  \rho_3^\dagger (t)\rho_3 (t_s) \vert 0 \rangle
\end{array}
\right)
\end{equation}
has been constructed, where $\rho_3(t)$ is an interpolating operator for the neutral $\rho$ meson with the momentum ${\bf p}$ and the polarization vector parallel to ${\bf p}$, and $(\pi\pi)(t)$ is an interpolating operator for two  pions, $(\pi\pi)(t) =(\pi^{-}({\bf p},t)\pi^{+}({\bf 0},t)-\pi^{+}({\bf p},t)\pi^{-}({\bf 0},t) )/\sqrt{2}$.  Two energy levels are obtained from two eigenvalues $\lambda_n(t,t_R)$ of the matrix $M(t,t_R) = G(t) G^{-1}(t_R)$ as
\begin{equation}
\lambda_n(t,t_R) = e^{- W_n(t-t_R)}
\end{equation}
for large $t$. To keep symmetry between source and sink in $G(t)$ U(1) noises are introduced.
The total number of quark propagators per configuration becomes 520, including 10 noises times 2 source points\cite{Ishizuka0}. 

\begin{figure}[tb]
\centering
\includegraphics[width=70mm,clip]{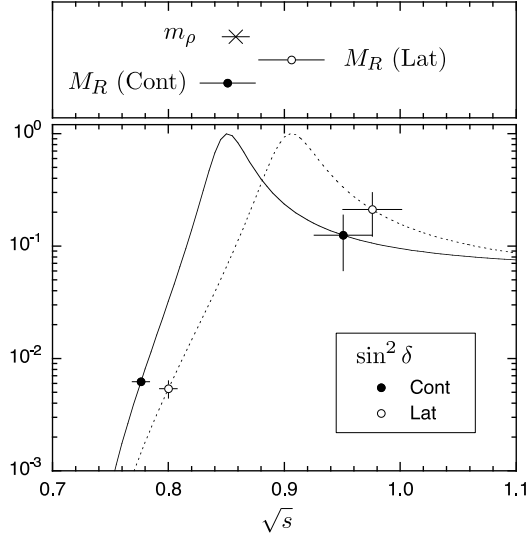}
\caption{Scattering phase shift $\sin^2\delta$(lower panel), positions of $m_\rho$ and resonance mass $M_R$ (upper panel) in lattice units. "Cont" refers to results obtained with the continuum dispersion relation while "Lat" to those obtained with the lattice dispersion relation.}
\label{fig:rho_width}
\end{figure}

From the energy levels $W_n$ extracted,
the invariant mass and the momentum $k$ are given by $\sqrt{s}=\sqrt{W_n^2-p^2}$ and $k^2=s/4 - m_\pi^2$, assuming the continuum dispersion relation, while they become
$\cosh(\sqrt{s}) =\cosh(W_n) - 2\sin^2(p/2)$ and $2\sin^2(k/2) = \cosh(\sqrt{s}/2) - \cosh(m_\pi)$,
using a lattice dispersion relation.
Then $W_1$ leads to $\tan\delta = 0.0791(9)$ at $\sqrt{s}=0.776(8)$ (continuum dispersion) or
$\tan\delta = 0.074(7)$ at $\sqrt{s}=0.800(8)$ (lattice dispersion), and
$W_2$ gives $\tan\delta = -0.38(11)$ at $\sqrt{s}=0.95(3)$ (continuum) or
$\tan\delta = -0.52(14)$ at $\sqrt{s}=0.98(3)$ (lattice).
In both cases
it is observed that $\delta > 0$ (attractive) at $\sqrt{s} < m_\rho=0.86(1)$ while $\delta < 0$ (repulsive) at $\sqrt{s} > m_\rho$. This property confirms the existence of a resonance at a mass around $m_\rho$\cite{SY}.

In Fig.~\ref{fig:rho_width}, $\sin^2\delta$, which is proportional to the scattering cross section of the two-pion system, is plotted against the invariant mass $\sqrt{s}$.
The two lines in the figure represent fits of the  data by
\begin{equation}
\tan \delta = \frac{g_{\rho\pi\pi}}{6\pi}\frac{k^3}{\sqrt{s}(M_R^2-s)},
\end{equation}
where $M_R$ is the resonance mass, plotted also in the upper panel of Fig.~\ref{fig:rho_width},
and $g_{\rho\pi\pi}$ is the effective coupling of $\rho\pi\pi$.
If $g_{\rho\pi\pi}$ were obtained at several quark masses, it could be extrapolated to the physical quark mass. Since this calculation is performed only at one quark mass, however,
it is assumed that $g_{\rho\pi\pi}$ is independent of the quark mass, so that the $\rho$ meson decay width at the physical quark mass is estimated by
\begin{equation}
\Gamma^{\rm ph} = \frac{g_{\rho\pi\pi}^2}{6\pi} \frac{(k^{\rm ph})^3}{(m_\rho^{\rm ph})^2}
=g_{\rho\pi\pi}^2\times 4.128 {\rm MeV},
\end{equation}
where $(k^{\rm ph})^2 = (m_\rho^{\rm ph})^2/4- (m_\pi^{\rm ph})^2$.
The fit result of $g_{\rho\pi\pi}$ gives $\Gamma^{\rm ph} = 162(35)$ MeV (continuum) and
$\Gamma^{\rm ph} = 140(27)$MeV (lattice), which are consistent with the experiment value, $\Gamma = 150$ MeV.

Since the first attempt is successful, the next step is the chiral extrapolation. The $\rho$ meson decay width is a good bench mark quantity which shows the dynamical nature of QCD vividly. Therefore  collaboration groups which have full QCD configurations at small quark masses are encouraged to calculate this quantity.
 
\section{Potentials for heavy hadrons}
\begin{figure}[b]
\centering
\includegraphics[width=45mm,clip]{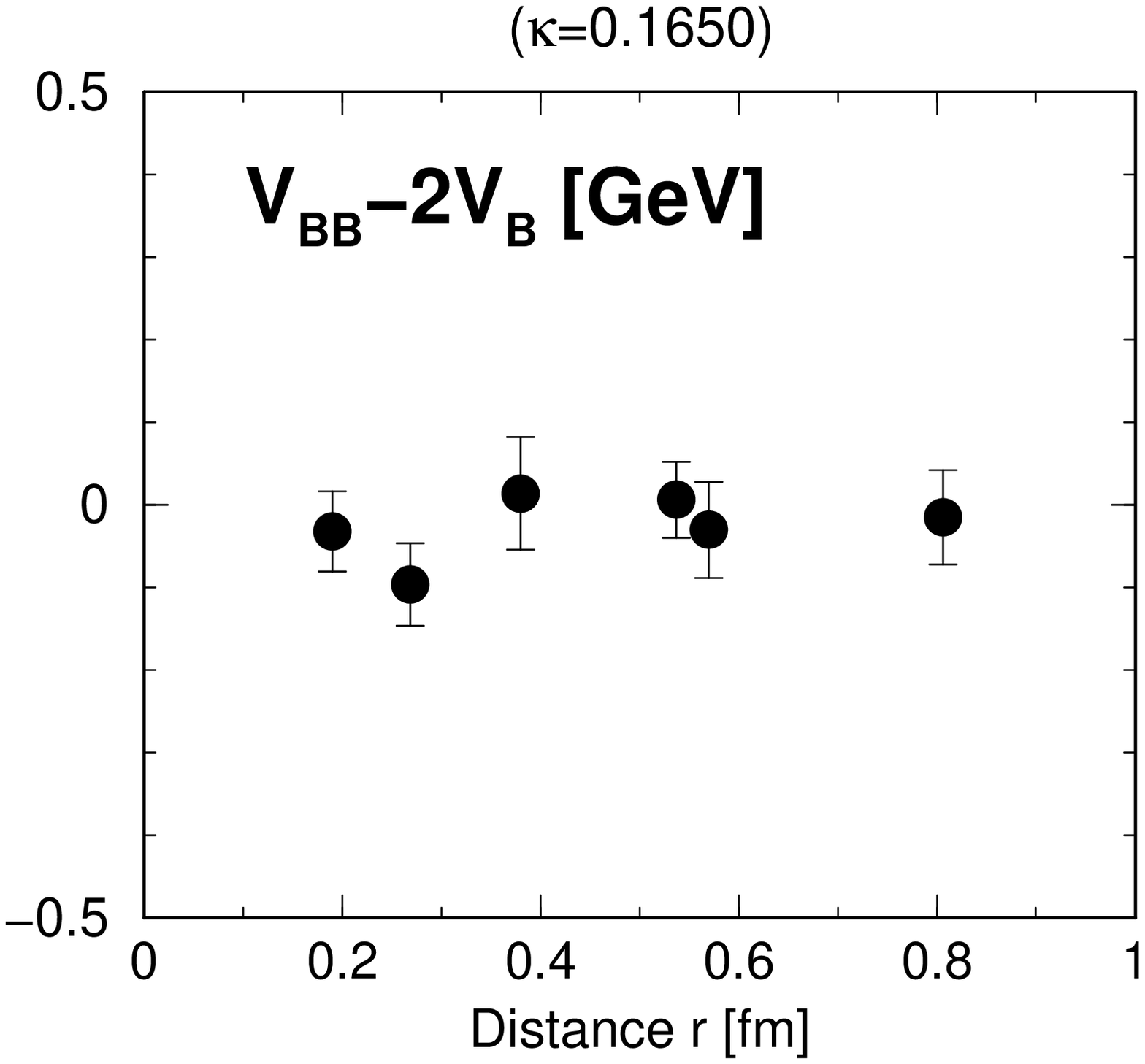}
\hspace*{2mm}
\includegraphics[width=45mm,clip]{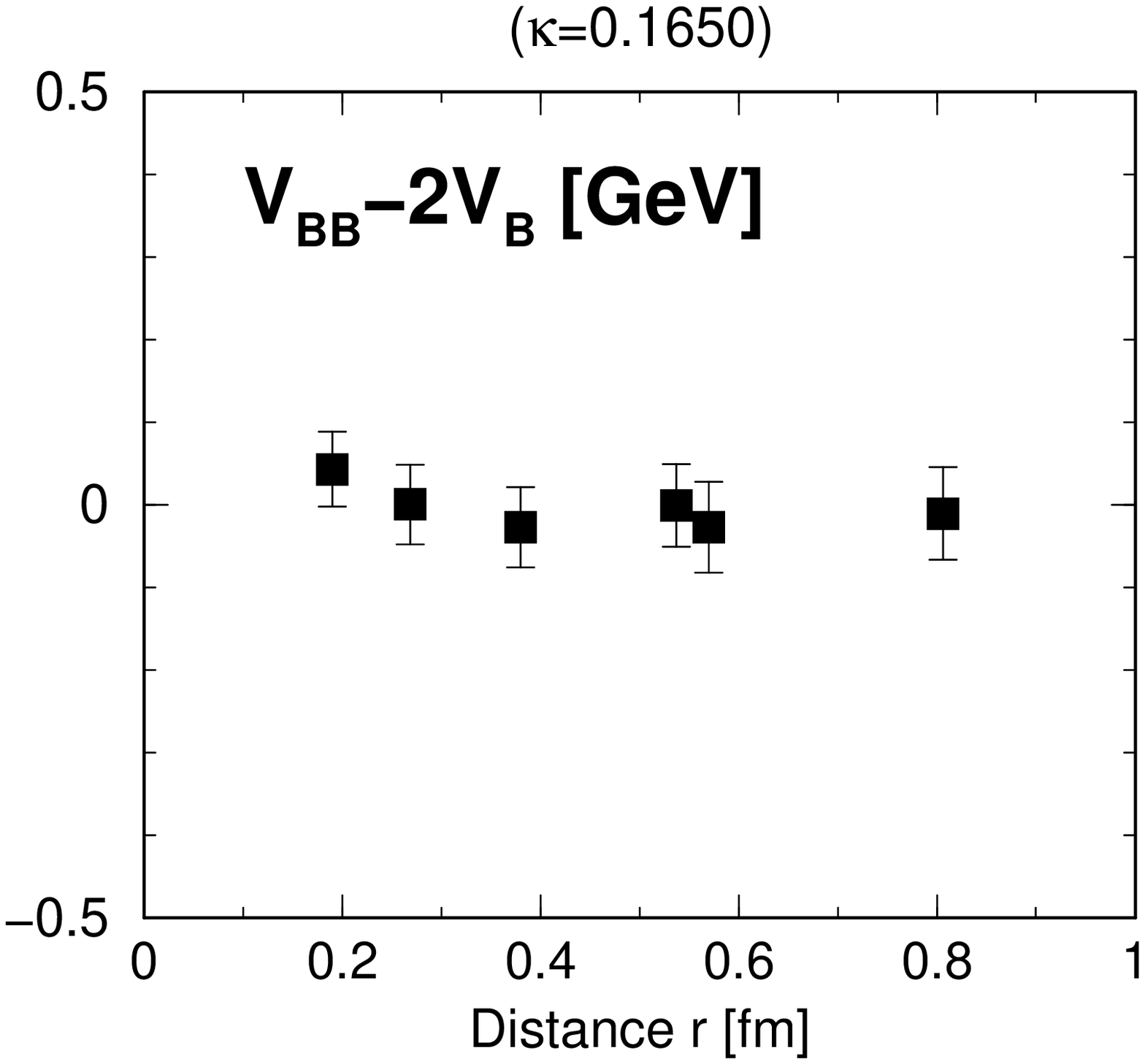}
\hspace*{2mm}
\includegraphics[width=45mm,clip]{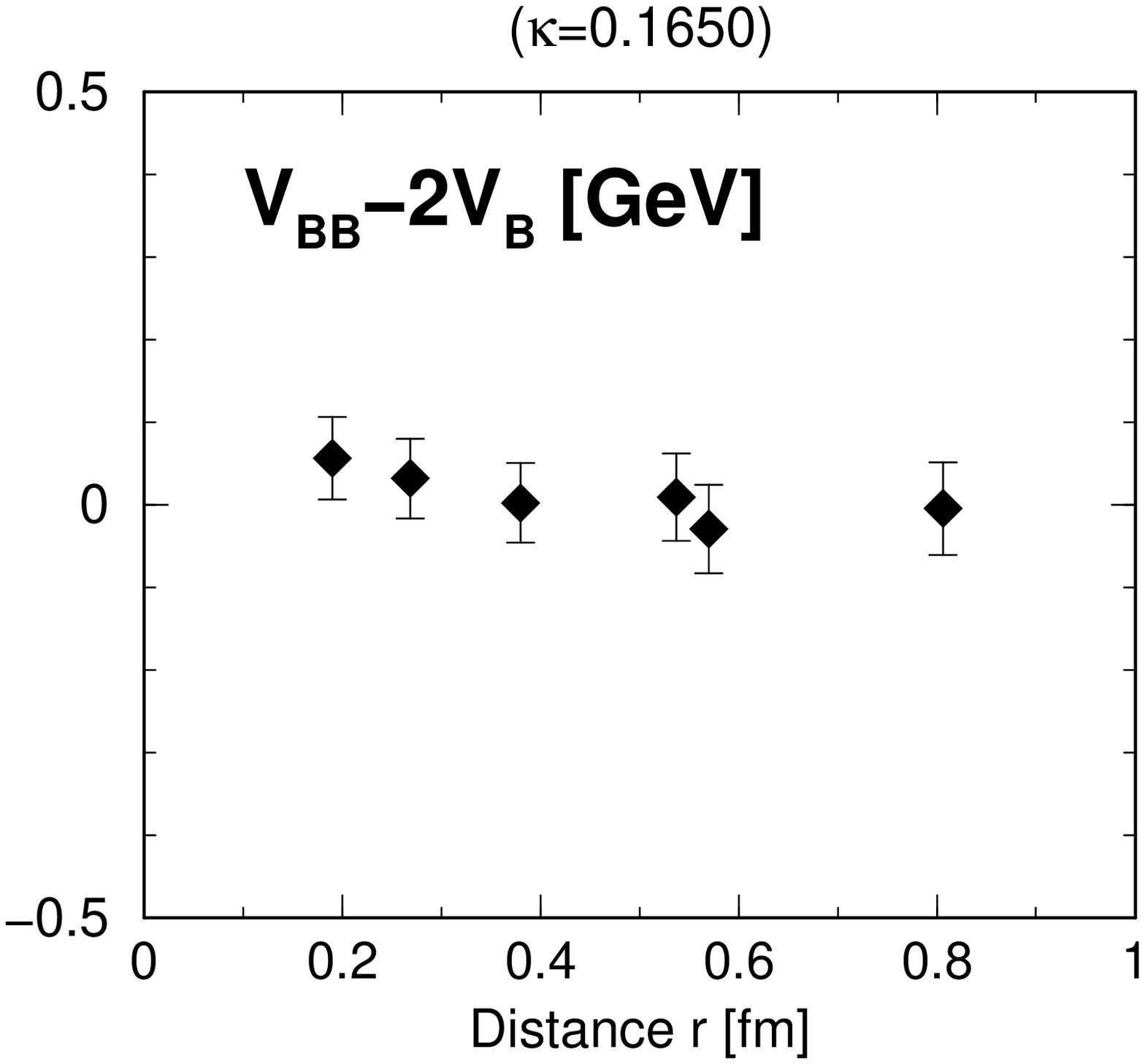}
\caption{The binding energy, $V_{\rm BB}- 2V_{\rm B}$ (GeV),  as a function of $r$ (fm).
Left: All the light-quark flavors are different. Middle: One pair of quark flavors is identical.
Right: Two pairs are identical.}
\label{fig:heavy_baryon}
\end{figure}

In the following two sections, extractions of the potential between hadrons from lattice QCD are discussed.
A straight-forward way to define a potential is to calculate an energy of a two-hadron system as a function of the distance between the two hadrons. A difficulty exists in the definition of the distance between two hadrons, since the two hadrons are moving around  changing their relative distance. 
To overcome this difficulty, one (infinitely) heavy quark may be introduced in each hadron, so that the distance between two hadrons is defined by the distance between the two heavy static quarks, which do not move in space for all the time. This definition, similar to the static quark potential from a Wilson loop or Polyakov lines, is indeed employed to calculate potentials between two heavy hadrons in lattice QCD.

\subsection{Baryon-Baryon}  
In Ref.~\cite{TDS1}, the energy of two heavy baryons has been measured as a function of the distance $r$ between two heavy quarks $Q$, using quenched QCD with the plaquette gauge action and the Wilson quark action 
at $a\simeq 0.19$ fm on a $20^3\times 24$ lattice. The binding energy of  two baryons, $V_{\rm BB}- 2 V_{\rm B}$,  is plotted as a function of $r$ in Fig.~\ref{fig:heavy_baryon} at $m_\pi\simeq 500$ MeV for the light quark,
where $V_{\rm BB}$ is an energy for two baryons while $V_{\rm B}$ is an energy for one baryon.
The differences of the three figures are explained in the caption. Surprisingly, the binding energy is very small and shows almost no dependence on $r$ for all three cases.  Thus, a repulsive core 
is not seen for the heavy baryon potential in this calculation.

\subsection{Meson-Meson and others}
\begin{figure}[tb]
\centering
\includegraphics[width=140mm,clip]{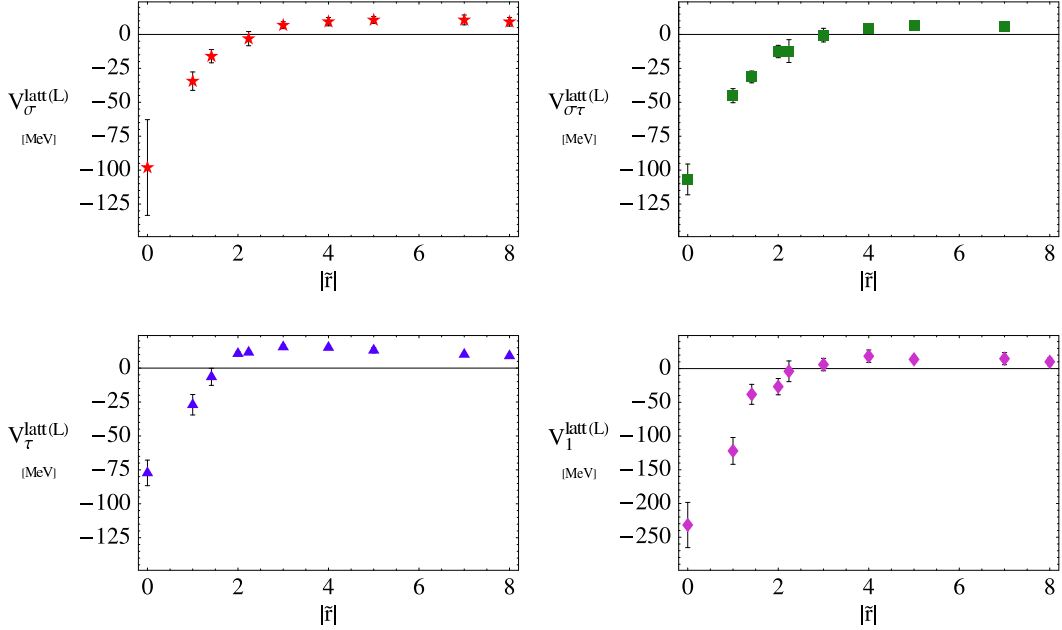}
\caption{The central meson-meson potential (MeV) as a function of $r$ in lattice units.}
\label{fig:BB_potential}
\end{figure}

In Ref.~\cite{NPLQCD3}, the binding energy of  two heavy-light mesons has been computed, each of which is made of one static quark and one light quark, using quenched QCD with the DBW2 gauge action at $a\simeq 0.1$ fm and the Wilson quark action at $m_\pi \simeq 400$ MeV. 
The central potential for total spin $S$ and isospin $I$ of two light quarks in a meson-meson system is decomposed as
\begin{equation}
V_{I,S}( r) = V_1(r) + (\sigma_1\cdot\sigma_2) V_{\sigma}(r) +(\tau_1\cdot\tau_2) V_{\tau}(r)+
(\sigma_1\cdot\sigma_2) (\tau_1\cdot\tau_2) V_{\sigma\tau}(r)
\end{equation}
where 
$\sigma_i$ ($\tau_i$) acts on the spin (isospin) of the light quark $q_i$, and $r$ is the distance between the two static quarks.
Results for $V_X(r)$ ($ X=1,\sigma,\tau,\sigma\tau$) are plotted in Fig.~\ref{fig:BB_potential}.
All potentials show attraction at short distance ($ r\le 0.2$ fm ).

\begin{figure}[bt]
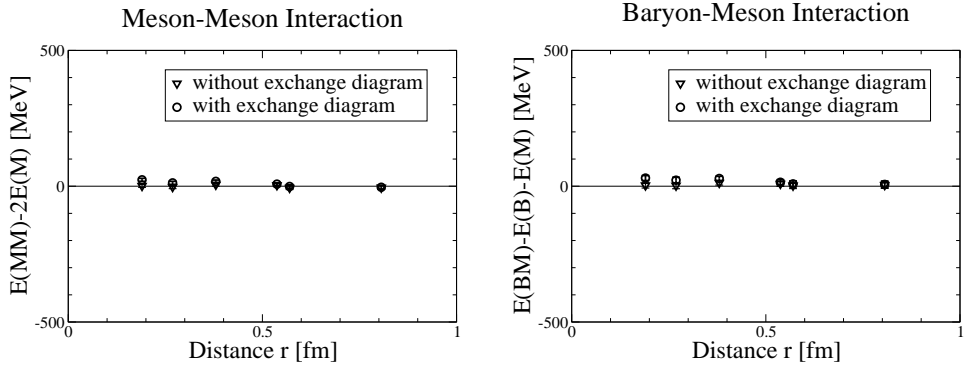

\centering
\includegraphics[width=60mm,clip]{Fig/MM_potential.eps}
\hspace*{5mm}
\includegraphics[width=60mm,clip]{Fig/MB_potential.eps}
\caption{The meson-meson potential (left) and the meson-baryon potential as a function of $r$ at $m_\pi\simeq 700$ MeV.  Triangles (circles) denote results without (with) light quark exchange diagrams. }
\label{fig:MM_MB_potential}
\end{figure}

In Ref.~\cite{TDS2}, a meson-meson potential and a meson-baryon potential has been calculated
with the same setup of the baryon-baryon case in the previous subsection. As shown in Fig.~\ref{fig:MM_MB_potential}, however, the dependence on $r$ is very small in both cases. In particular, no short-distance attraction is observed for the two heavy-light mesons, contrary to the results in Fig.~\ref{fig:BB_potential}. 
There are several candidates which might cause the difference: the light quark mass, the lattice spacing, the lattice volume, the statistics or the way to extract the binding energy. Further investigations are necessary for definite conclusions about the static quark approach to potentials between hadrons. 

\section{Potentials from wave functions}
In Ref.\cite{IAH1}, the nucleon-nucleon ($NN$) potential has been extracted in lattice QCD by a totally  new and different method, which is explained in this section.

\subsection{Wave functions and potentials}
The starting point of the new method for the $NN$ potential is a wave function defined by
\begin{eqnarray}
\varphi({\bf r}, E) &=& \langle 0 \vert N({\bf x},0) N({\bf y},0)\vert 2N; E\rangle,
\qquad {\bf r} ={\bf x} -{\bf y},
\label{eq:wave}
\end{eqnarray}
where $N({\bf x}, t)$ is an interpolating field of the nucleon, $\vert 2N;E\rangle $ is a 2$N$ state with energy $E < E_{\rm inelastic}$ with the inelastic threshold energy $E_{\rm inelastic}$ of the 2$N$ system.
This wave function was first analyzed for a spin model in Ref.~\cite{spin}, and has been employed in lattice QCD to calculate the $\pi\pi$ scattering length\cite{Ishizuka1}.

This wave function has the following properties.
For large enough $r=\vert {\bf r}\vert$,  $(H_0 - E) \varphi({\bf r},E) = 0$,  where $H_0 = \displaystyle\frac{- \nabla^2}{2\mu}$ is a free non-relativistic Hamiltonian
with the reduced mass $\mu = m_N/2$. This means that the interaction between two nucleons vanishes for a large separation. Moreover, for example in the case of the $^1S_0$ channel, it can be shown that
\begin{eqnarray}
\varphi^S({\bf r},E) &\sim& e^{i\delta_0(k)} \frac{\sin (k r +\delta_0(k))}{kr} + \cdots,
\label{eq:asymptotic}
\end{eqnarray}
where $\delta_0(k)$ is the $S$ wave scattering phase shift and $k$ is determined by $E = k^2/(2\mu)$.
 In Ref.~\cite{Ishizuka1,Lin1}, it is shown for a two-pion system that
\begin{eqnarray}
\varphi^S({\bf r},E) &=& e^{i{\bf k r}} + \frac{i k}{4\pi r}H(k;k)j_0(kr) + {\bf P}\int \frac{d^3p}{(2\pi)^3}\frac{H(p;k) j_0(pr)}{p^2-k^2} + I_{\rm others}({\bf r}),
\label{eq:exact}
\end{eqnarray}
where $H(p,k) =  (E_k+ E_p) M(p,k)/(8E_k E_p)$ with $E_p=\sqrt{m_\pi^2+p^2}$ and $M(p,k)$ is the (off-shell) scattering amplitude.
A similar but a little more complicated expression can be shown also for the $NN$ system\cite{AHI}.
For large $r$, if  $E < E_{\rm inelastic}$, the contribution from $I_{\rm others}$ vanishes exponentially and eq.(\ref{eq:exact}) approaches eq.(\ref{eq:asymptotic}).  

The potential is defined from the wave function through the Schr\"odinger equation
\begin{eqnarray}
[H_0 + V({\bf r}) ]\varphi ({\bf r},E) &=& E \varphi({\bf r},E),
\label{eq:def_potential0}
\end{eqnarray}
which symbolically gives 
\begin{eqnarray}
V({\bf r}) &=& \frac{ (E-H_0)\varphi({\bf r},E)}{\varphi({\bf r},E)} .
\label{eq:def_potential}
\end{eqnarray}

\subsection{Lattice calculation}
A  corresponding lattice wave function in a finite box at $E\simeq 0$ is given explicitly by
\begin{eqnarray}
\varphi({\bf r},E) &\equiv&\frac{1}{24}\sum_{{\cal R}\in {\cal O}}
\frac{1}{L^3}\sum_{\bf x} P_{ij}^I P_{\alpha\beta}^S \langle 0 \vert N_\alpha^i({\cal R}[{\bf r}]+{\bf x},0) N({\bf x},0)\vert 2N; E\rangle,
\end{eqnarray}
where $N_\alpha^i = \varepsilon_{abc}(^t q^a C\gamma_5\tau_2 q^b) q_{\alpha}^{i,c}$,
$P^1=1$ and $P^0 =\tau_2$ are isospin projections, $P^1=1$ and $P^0 =\sigma_2$ are spin projections, and the summation over ${\bf x}$  gives zero total momentum. The summation over the discrete rotations ${\cal R}$ of the cubic group ${\cal O}$ implies that the state belongs to the $A_1^+$ representation of the cubic group, which is expected to couple to an $L=0$ ground state as well as $L\ge 4$ excited states of the rotation group in the continuum theory.

The wave function without projections has been extracted from the 4-point nucleon correlator as
\begin{eqnarray}
F_{\rm NN}({\bf x},{\bf y}, t; t_0) &\equiv& \langle 0 \vert N_{\alpha}^i({\bf x}, t) N_\beta^j({\bf y},t){\cal J}_{\rm NN}\vert 0\rangle
=\sum_n A_n \langle 0 \vert N_{\alpha}^i({\bf x}, t) N_\beta^j({\bf y},t) \vert 2N; E_n\rangle e^{i E_n(t-t_0)}
\end{eqnarray}
where $A_n = \langle 2N; E_n \vert {\cal J}_{\rm NN}(t_0)\vert 0\rangle$ and ${\cal J}_{\rm NN}(t_0)
= P_{ij}^I P_{\alpha\beta}^S {\cal N}_\alpha^i{\cal N}_\beta^j $. In ${\cal N}$, a wall source 
is employed by replacing $q({\bf x},t_0)$ with $Q(t_0) =\sum_{\bf x} q({\bf x},t_0)$ after Coulomb
gauge fixing.  For large enough $t$ the wave function for $E=E_0$ is obtained.

In a finite volume, energy levels are shifted from those in the infinite volume as
$\Delta E_n (L) = E_n(L) - E_n(L=\infty) =O(L^{-3})$ due to interactions. From these shifts, the scattering phase shift can be extracted\cite{Luescher1}.

\subsection{Numerical simulations}
In the actual calculation, the NN scattering for $L=0$ is considered.
There are two channels, the spin singlet ($S=0$) channel $^1S_0$ and the spint triplet ($S=1$) channel $^3S_1$, where the standard notation $^{2S+1}L_J$ is used.  
Since only a central potential   appears  for $^1S_0$,  the definition (\ref{eq:def_potential}) directly gives $V_{\rm NN}({\bf r}) = V_{\rm C}(r)$. On the other hand, in the case of $^3S_1$, the potential becomes
$V_{\rm NN}({\bf r}) = V_{\rm C}(r) + V_{\rm T}(r) S_{12}({\bf r})$, where $S_{12}({\bf r}) = 3({\bf\sigma}_1\cdot{\bf r})({\bf\sigma}_2\cdot{\bf r}) -({\bf\sigma}_1\cdot{\bf\sigma}_2)$ is the tensor operator. The tensor potential induces mixing between $^3S_1$ and $^3D_1$ states, so that
the definition (\ref{eq:def_potential}) gives the so-called effective central potential, which includes the mixing effect from the $^3D_1$ state as a second order perturbation.

Numerical simulations have been made in quenched QCD on a $32^4$ lattice with the plaquette gauge action and the Wilson quark action, at $a\simeq 0.137$ fm from the $\rho$ meson mass.  The number of configurations is 2000 for  $m_\pi \simeq 370$ MeV and 527 MeV, and 1000 for $m_\pi \simeq$ 732 MeV. A Dirichlet boundary condition (DBC) in time and a periodic B.C.  in space are employed, and the wall source with Coulomb gauge fixing is placed at $t_0=5$, to avoid an influence
of the  DBC.  Calculations have been performed on a Blue Gene/L at KEK, which has 57.3 TFlops peak performance. It took about 4000 hours of 512 Nodes (a half-rack, 2.87TFlops peak) with
34--48\% sustained speed to complete them.

\subsection{Results}
\begin{figure}[bt]
\centering
\includegraphics[width=52mm,angle=270, clip]{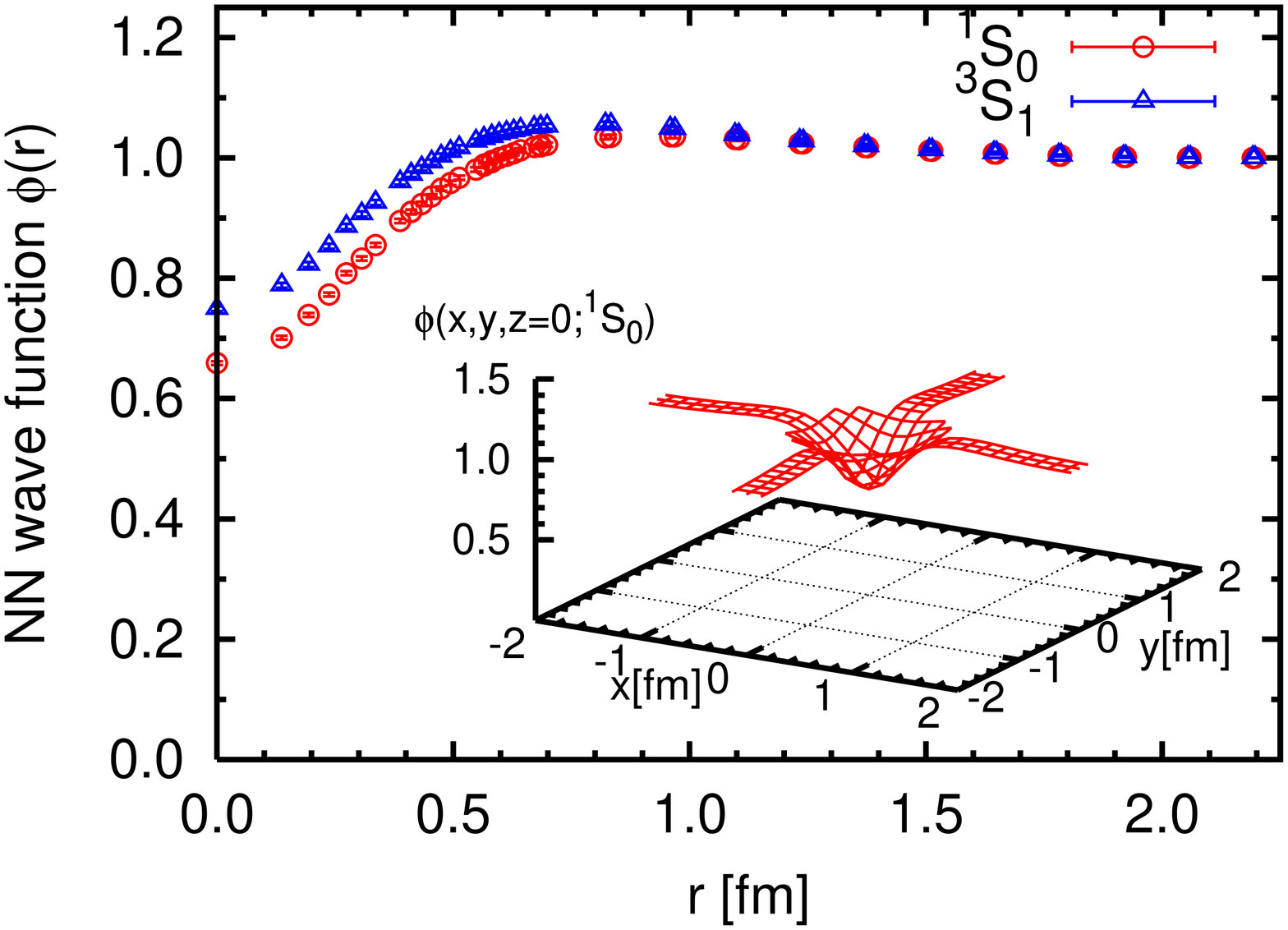}
\includegraphics[width=52mm,angle=270, clip]{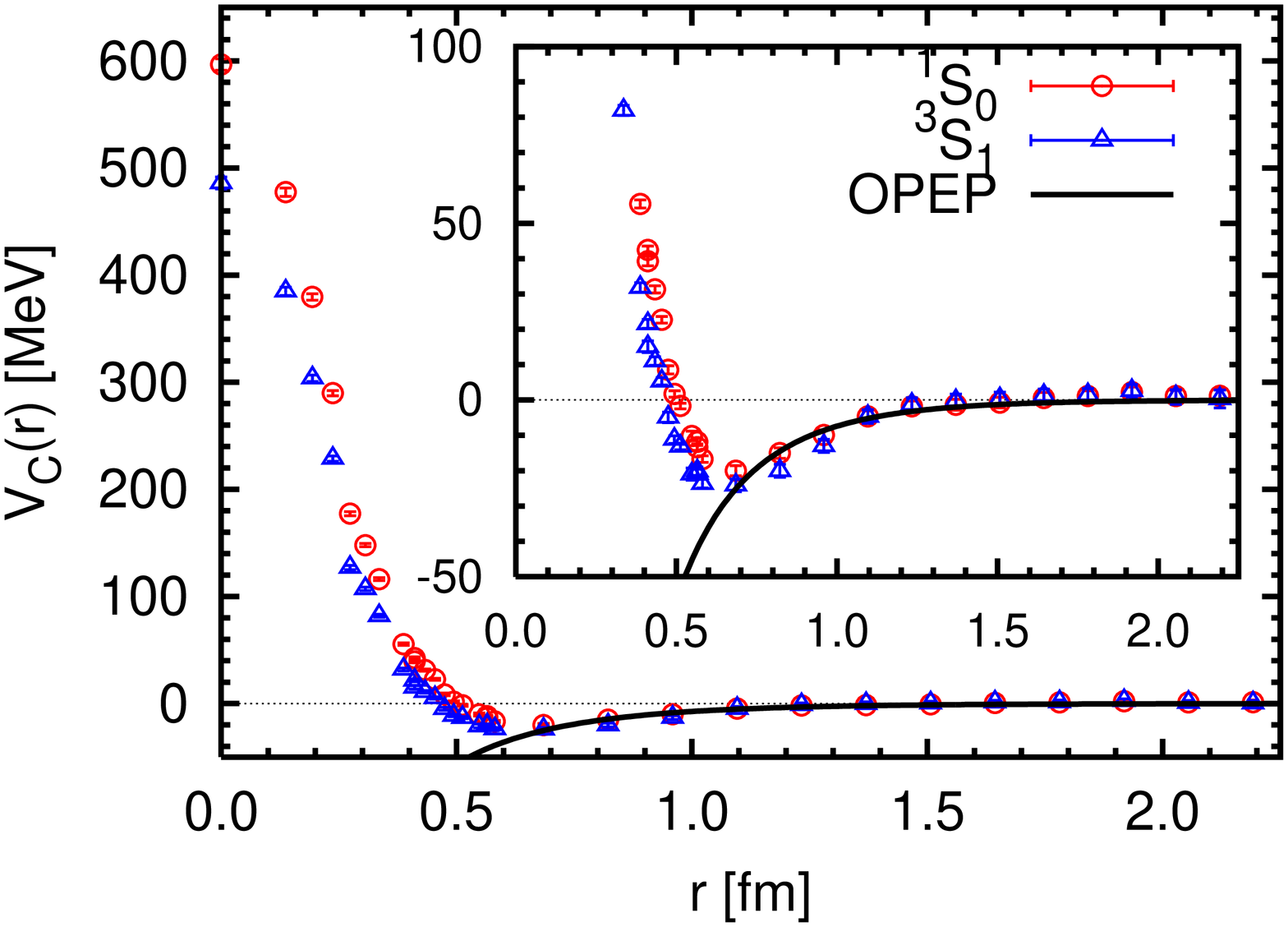}
\caption{Figures from \cite{IAH1}. Left: The normalized $NN$ wave function for the singlet(circles) and the triplet(triangles) as a function of $r$ (fm) at $m_\pi\simeq 527$ MeV.
Right: The central (effective central) $NN$ potential (MeV) as a function of $r$ (fm) for the singlet (triplet) at $m_\pi\simeq 527$ MeV. }
\label{fig:wave_function}
\end{figure}

In the left panel of Fig.~\ref{fig:wave_function}, the $NN$ wave function normalized to 1 at $r\simeq 2.2$ fm
is plotted as a function of $r$ for $^1S_0$ and $^3S_1$ at $m_\pi \simeq $ 527 MeV. In both cases the wave function shows an increase at 0.5 fm $ < r < $ 1.5 fm, suggesting attraction, while it decreases at $ r < 0.5$ fm, indicating the existence of the repulsive core. In the right panel of Fig.~\ref{fig:wave_function} the central (effective central) $NN$ potential extracted from the wave function by eq.(\ref{eq:def_potential}) at $m_\pi\simeq 527$ MeV is plotted as a function of $r$ for $^1S_0$ ($^3S_1$). Interestingly the potential obtained qualitatively agrees with the $NN$ potential determined from scattering experiments: weak attraction at long distance, a little stronger attraction at intermediate distance and   strong repulsion at short distance (the repulsive core). The solid line is the Yukawa potential (One Pion Exchange Potential) given by
\begin{equation}
V_{\rm C}^{\rm Yukawa}(r) =\frac{ g_{\pi N}^2}{4\pi} \frac{({\bf\tau}_1\cdot{\bf\tau}_2) ({\bf\sigma}_1\cdot{\bf\sigma}_2)}{3}\left(\frac{m_\pi}{2m_N}\right)^2 \frac{e^{-m_\pi r}}{r},
\end{equation}
which agrees well with the data at long distance,
with $g^2_{\pi N}/(4\pi) \simeq 14.0$ from experiments, $m_\pi \simeq 0.527$ GeV and $m_N\simeq 1.34$ GeV from lattice data.  Thus, in some sense, the Yukawa theory for the nuclear force at
$ r > 1$ fm is confirmed by lattice QCD.

\begin{figure}[bt]
\centering
\includegraphics[width=70mm,angle=270, clip]{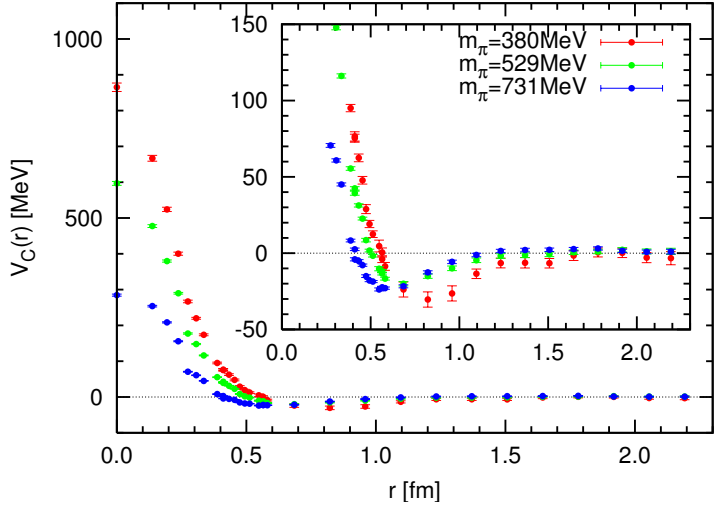}
\caption{The central  NN potential (MeV) as a function of $r$ (fm)
for $^1S_0$ at $m_\pi\simeq 320$(red), 527(green) and 732(blue) MeV \cite{IAH2}.}
\label{fig:potential_qmass}
\end{figure}

The quark mass dependence of the $NN$ potential for $^1S_0$ is shown in Fig.~\ref{fig:potential_qmass}.
As the quark mass decreases, the repulsive core at short distance gets stronger, and at the same time,  
attraction at intermediate distances also becomes a little stronger\cite{IAH2}.
In the future it will be interesting to see if this quark mass dependence remains in full QCD.

\section{Theoretical considerations and future perspectives}

\subsection{Energy dependence of the potential}
Since the wave function depends on the energy $E={\bf p}^2/(2\mu)$, the potential defined by eq.(\ref{eq:def_potential}) is also energy dependent in general:
\begin{equation}
V^J( r,E) = \frac{(E-H_0)\varphi^J(r,E)}{\varphi^J( r,E)} ,
\end{equation}
where $J$, the total angular momentum, is fixed, and $r=\vert{\bf r}\vert$. To make our argument simpler, the spin degrees of freedom are not considered here.
Since $V^J( r,E)$ carries more information than the scattering phase shift $\delta (E)$ does, $V^J( r,E)$ is redundant and therefore not physical. 
For the potential defined from the wave function to be physically meaningful, its energy dependence must be weak for some range of small $E$. 

\begin{figure}[bt]
\centering
\includegraphics[width=65mm, clip]{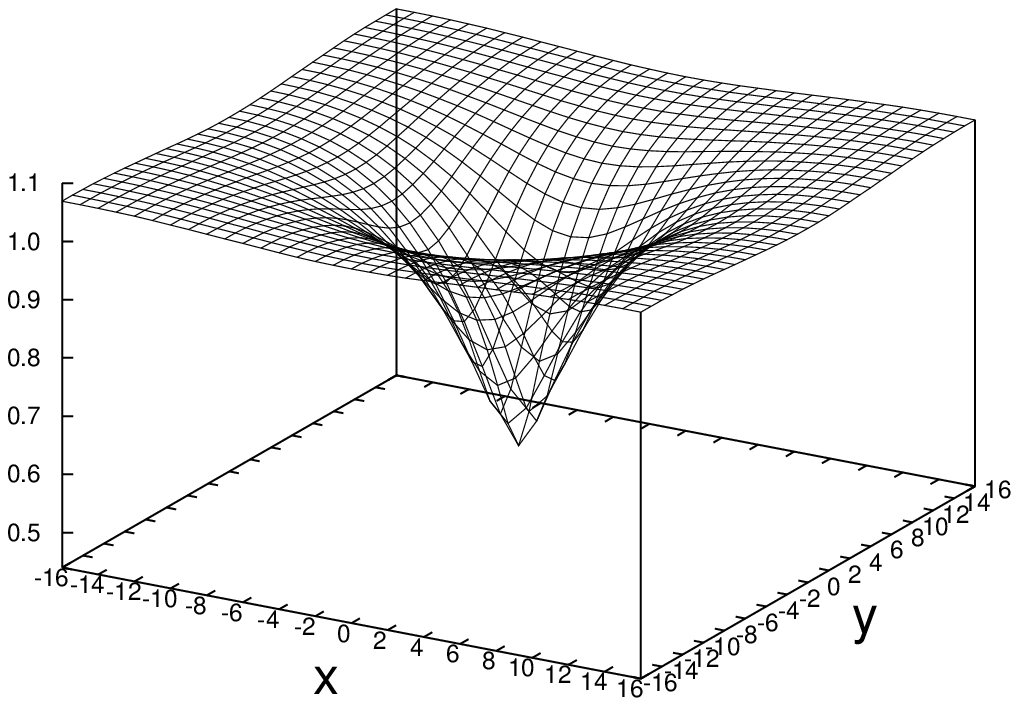}
\includegraphics[width=65mm, clip]{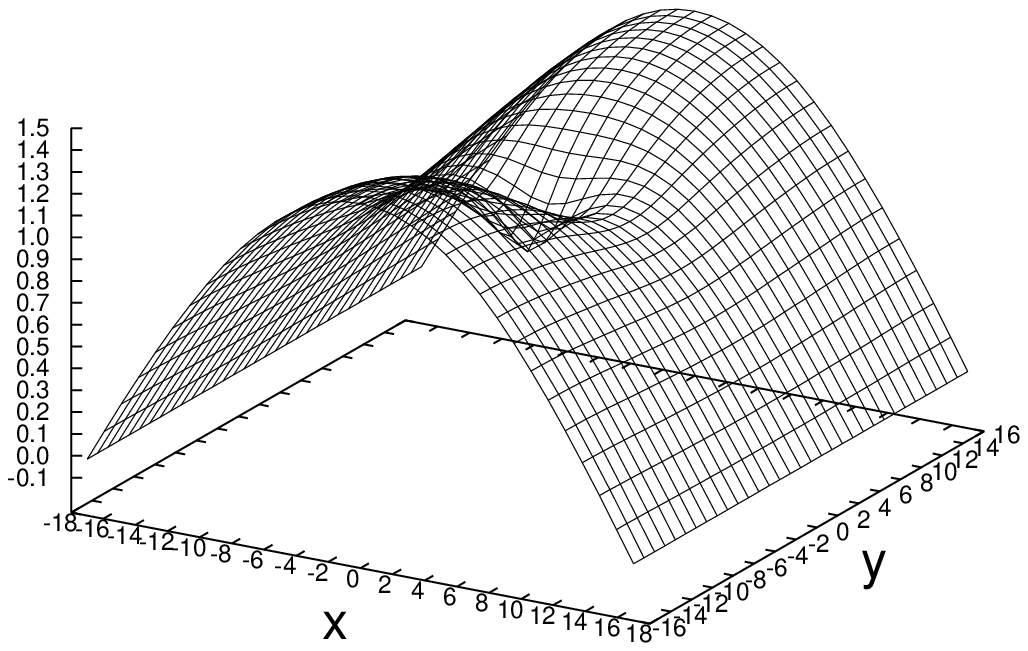}
\includegraphics[width=65mm, clip]{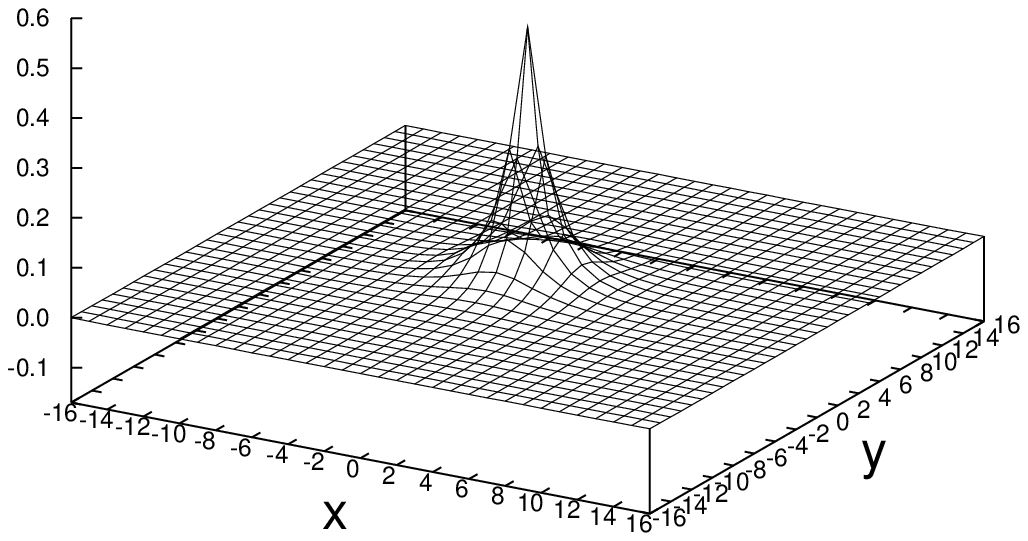}
\includegraphics[width=65mm, clip]{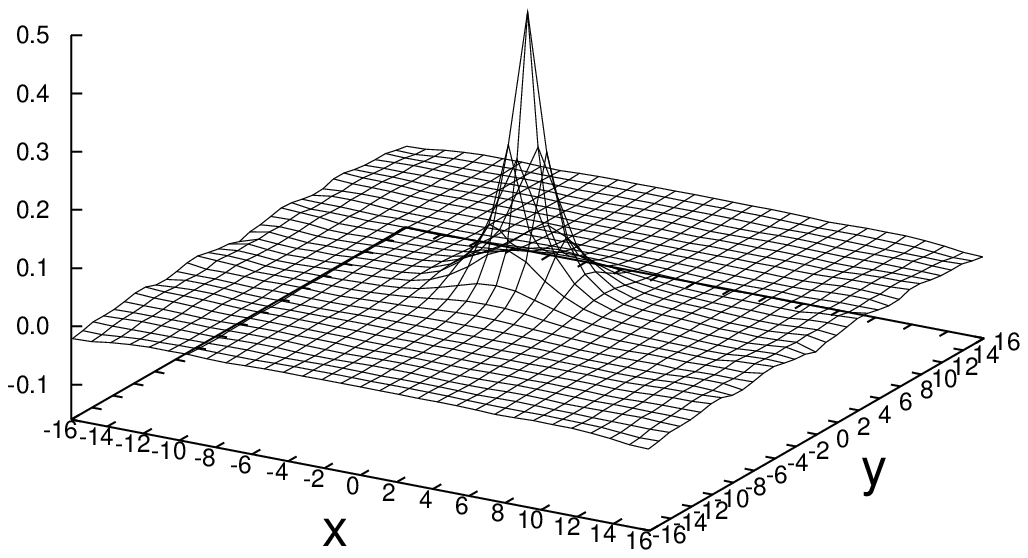}
\caption{Upper plots: The $\pi\pi$ wave function at $p_x=0, t=0$ (left) and $p_x=2\pi/L, t= v\gamma x$ (right)
in quenched QCD at $m_\pi/m_\rho \simeq 0.51$.
Lower plots: The corresponding potential.}
\label{fig:pipi_potential}
\end{figure}

In Fig.~\ref{fig:pipi_potential}, the $\pi\pi$ wave function and the corresponding potential in the $S$ channel are shown at ${\bf r}=(x, y, 0)$ in quenched QCD at $m_\pi/m_\rho \simeq 0.51$ for ${\bf p}={\bf 0}$ (left) and ${\bf p} =(2\pi/L,0,0)$ (right) on an $L^3$ box\cite{Sasaki} .
Although the wave functions at the two different energies are very different,
the potentials look similar. For the $\pi\pi$ system, the energy dependence of the potential is not so strong at low energy.
Note, however, that
the wave function (and therefore the potential) at ${\bf p} =(2\pi/L,0,0)$ is calculated at equal time in the laboratory system and
is transformed back to the center of mass system, so that the relative time between the two pions is $x$ dependent: $t = v \gamma x$.

\subsection{A unique local potential}
In this subsection a method to extract a unique potential from wave functions is proposed\cite{AHI}.
As mentioned in the previous subsection, the potential seems $E$ dependent in general. In addition, it may also depend on the choice of the interpolating operator $N({\bf x},t)$ in (\ref{eq:wave}):  one can use
a different $\tilde N({\bf x},t)$ unless it changes the asymptotic behavior (the phase shift).

To begin with,  it is argued that, by introducing a non-local potential, the energy dependence of the potential can be removed. The non-local potential is defined by the following equation,
\begin{eqnarray}
K( r,E) &\equiv & (E- H_0)\varphi ( r, E) = \sum_{ r'} U ( r,  r')
\varphi( r', E),
\end{eqnarray}
where all quantities are taken to be real. For notational simplicity, $\sum$ is used even for the case that a variable is continuous and the superscript $J$ for the angular momentum is omitted here.  
Since $\{\varphi( r,E) \}_E$ form a complete set, there exists an inverse such that
\begin{eqnarray}
\sum_E \varphi( r, E)\varphi^{-1}(E, r' ) &=&\delta( r- r'), \qquad
\varphi^{-1}(E, r) = \sum_{E'}\eta^{-1}_{E,E'}\varphi (r, E')
\end{eqnarray}
where $\eta^{-1}_{E,E'}$ is the inverse of $\eta_{E,E'} =\sum_{r}\varphi( r,E)\varphi( r,E')$. Using this we  obtain
\begin{equation}
U (r, r') =\sum_E K(r,E)\varphi^{-1}(E, r').
\label{eq:non-local}
\end{equation}

In the actual simulations, it is impossible to obtain $\varphi( r,E)$ for all $E$. If $\varphi( r,E)$'s are obtained for $E=E_0, E_1,\cdots , E_n$, we expand $U$ in terms of a derivative $d$ such that
\begin{eqnarray}
U(r,r') &=& \left[ U_0(r) + U_1(r) d + U_2(r) d^2 +\cdots\right]\delta(r-r') = \sum_{k=0}^{n} U_k(r)d^k\delta(r-r') .
\end{eqnarray}
The coefficient $U_k$ can be obtained as
\begin{eqnarray}
U_k(r)&=& \sum_{j=0}^n K(r, E_j)\phi_{j,k}^{-1}(r)
\end{eqnarray}
where $\phi_{j,k}^{-1}$ is the inverse of $\phi_{j,k} = d^k\phi(r,E_j)$. Note that $d$ becomes a difference on a lattice.

It is now argued that the non-local potential $U$ obtained above can be transformed to a local $E$ independent potential $V$.  The Schr\"odinger equations for them become
\begin{eqnarray}
(H_0 + U) \vert \varphi, E \rangle &=& E\vert \varphi, E\rangle, \quad
\langle { r}\vert U\vert { r'}\rangle = U({ r}, { r'}),\quad
\langle { r}\vert \varphi, E\rangle =\varphi({ r},E) \\
(H_0 + V) \vert \psi, E \rangle &=& E\vert \psi, E\rangle, \quad
\langle { r}\vert V\vert { r'}\rangle = V({ r})\delta({ r}- { r'}),\quad
\langle { r}\vert \psi, E\rangle =\psi({ r},E) ,
\label{eq:local}
\end{eqnarray}
where $H_0+U$ is not hermitian unless $U(r, r')=U(r', r)$, while $H_0+V$ is hermitian.
Note here that the eigenvalues $E$ in both equations must be equal  to ensure that both wave functions give the same scattering phase shift (and the bound state spectrum if any). 
Existence and uniqueness of $V$ are suggested by the inverse scattering theory, which tells us that a local potential for a fixed $J$ is uniquely reconstructed from the scattering phase shift for all $E$ and informations of possible bound states.
Since the eigenfunctions $\vert \psi , E\rangle$ of the hermitian operator $H_0+V$ satisfy
$\langle \psi,E\vert \psi,E'\rangle = C_E^2\delta_{E,E'}$, in contrast to 
$\langle \varphi,E\vert \varphi,E'\rangle = \eta_{E,E'}$, the transformation function
$\langle { r}\vert \Lambda \vert { r'}\rangle =\Lambda ({ r},{ r'})$ defined by
\begin{eqnarray}
\vert \psi, E \rangle &=&  \Lambda \vert \varphi,E\rangle,
\quad
\vert \varphi,E\rangle = \Lambda^{-1} \vert \psi,E\rangle ,
\end{eqnarray}
can be constructed as
\begin{eqnarray}
\Lambda  &=& \sum_{E,E'} \vert \psi, E\rangle \eta^{-1}_{E,E'} \langle \varphi, E' \vert ,\qquad
\Lambda^{-1} = \sum_{E} \vert \varphi, E\rangle \frac{1}{C_E^2} \langle \psi, E\vert .
\end{eqnarray}
The compatibility of non-local and local potentials leads to
\begin{eqnarray}
 \Lambda (H_0+ U)  \vert \varphi, E \rangle = E \Lambda \vert \varphi, E \rangle
 = 
 (H_0+V)  \Lambda  \vert \varphi, E \rangle ,
\end{eqnarray}
which, together with the completeness of  $\vert \varphi, E \rangle$, implies
\begin{eqnarray}
[ \Lambda, H_0 ]  +\Lambda U &=&  V \Lambda.
\label{eq:master}
\end{eqnarray}
If  the number of degrees of freedom for $E$ and $r$ is denoted as $\# E= \# r= N$ on a finite lattice, 
eq.(\ref{eq:master}) gives $N\times N$ constraints, while the unkown
$\Lambda$ and $V$ have $N\times N$ and $N$ components, respectively.
Therefore the determination of both $\Lambda$ and $V$ from eq.(\ref{eq:master})  seems an ill-posed problem. However, $N(=\# E)$ components of $\Lambda$ can be taken freely, since the eigenvalue equation (\ref{eq:local}) does not depend on $C_E$, the norm of the eigenfunction, for all $E$'s.
Using this freedom, eq.(\ref{eq:master}) is enough to determine $\Lambda$ and $V$.
Eq.(\ref{eq:master}) is the master equation, which determines the unique local energy independent potential $V$, once
a non-local potential $U$ is obtained from a wave function defined through
 a particular choice of operators.  This conceptually solves the uniqueness (or $E$ dependence) problem of "the potential from a wave function".
The results in the previous section correspond to the 0th order solution of this equation: $V = U_0$ and $\Lambda=1$.  

\subsection{Future perspectives} 
\begin{figure}[tb]
\centering
\includegraphics[width=70mm, clip]{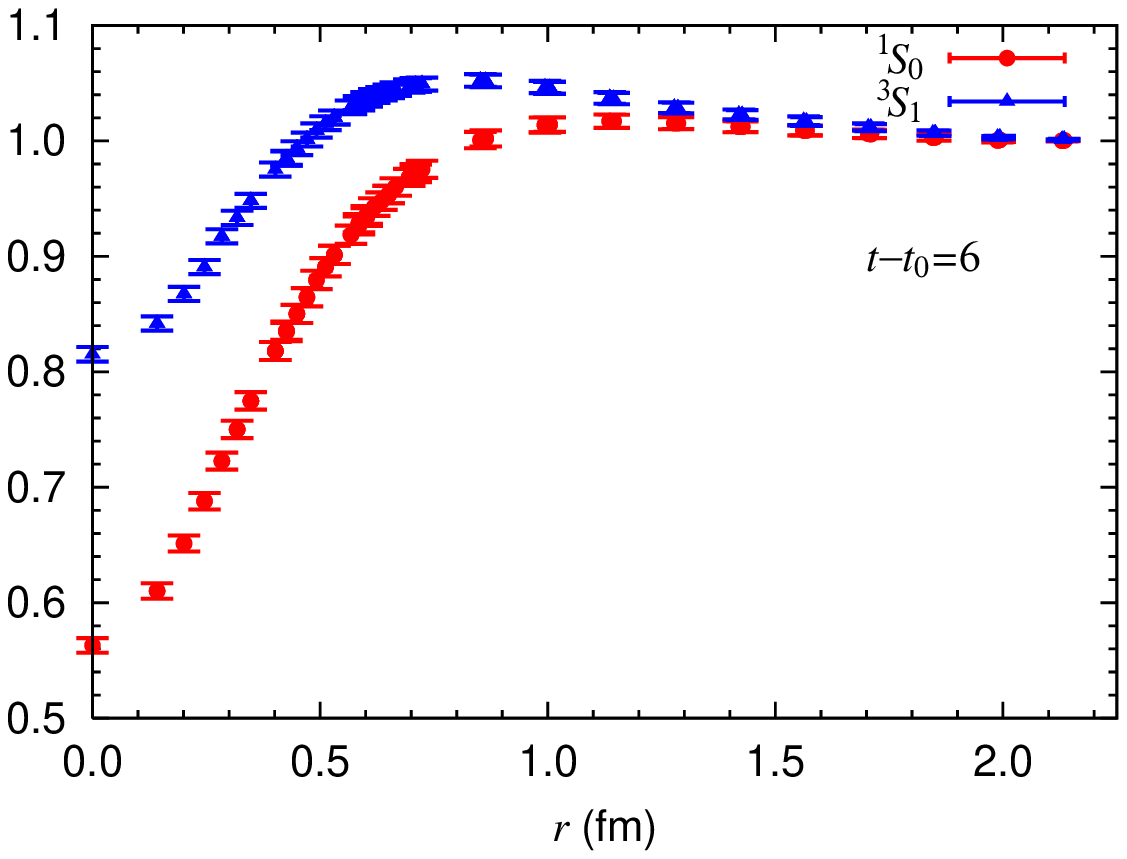}
\includegraphics[width=70mm, clip]{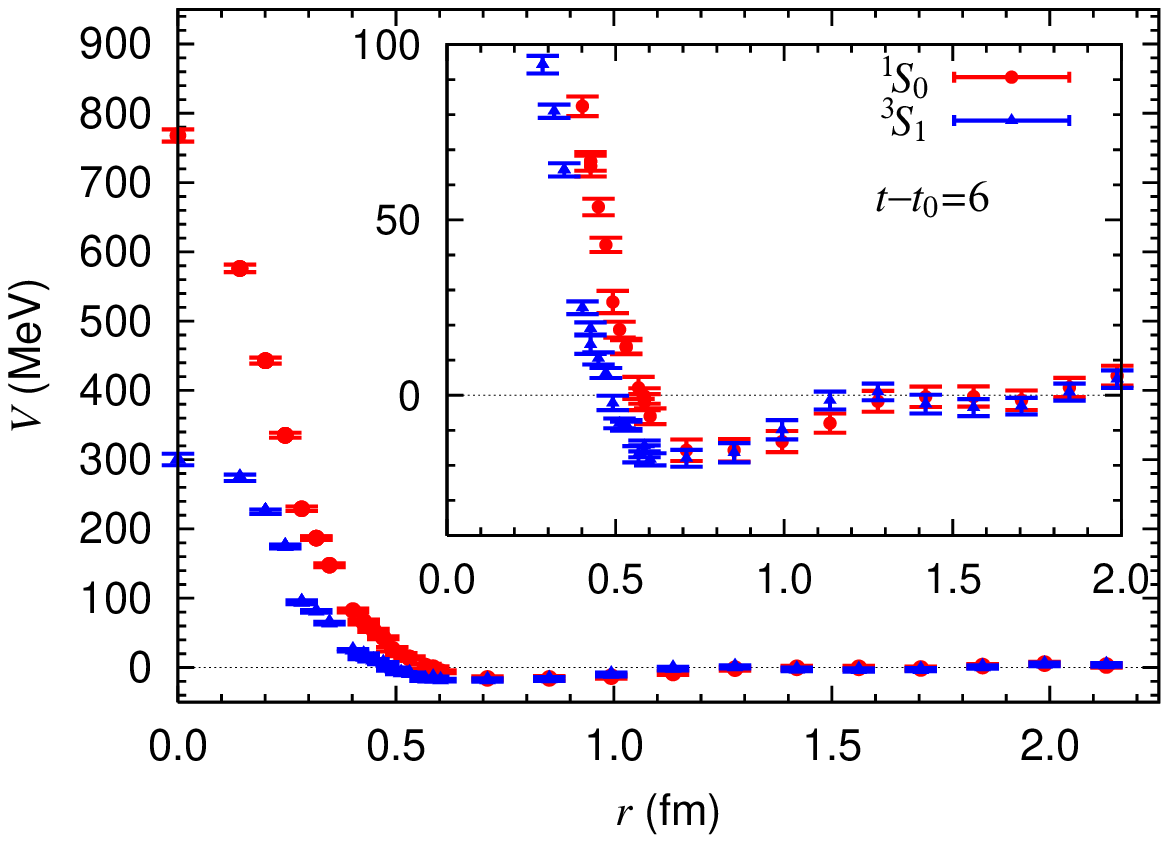}
\caption{Figures from \cite{NIAH}. Left: The  normalized wave function of $p \Xi^0$ in the $^1S_0$ (circles) and $^3S_1$ (triangles) channels as a function of $r$ (fm).
Right: The corresponding effective central potential (MeV) .}
\label{fig:YN_potential}
\end{figure}

To understand interaction properties of hyperons, baryons which include at least one strange quark,
is an important subjects in the nuclear physics.
Hyperon-nucleon ($YN$) and hyperon-hyperon ($YY$) interactions are relevant to structures of the neutron-star core and the existence/absence of H-dibaryon states.
Properties of hypernuclei, nuclei which contain hyperons, will be also studied,
as a project at J-PARC (Japan  Proton Accelerator Research Complex).
However, $YN$ and $YY$ interactions are poorly known  both theoretically and experimentally so far.
Lattice QCD calculations of the $NN$ potential in the previous section can be extended to $YN$ and $YY$ potentials.

In Fig.~\ref{fig:YN_potential}, the wave function and the corresponding potential for $p\Xi^0$ are plotted as a function of $r$ in quenched QCD\cite{NIAH}. The lattice parameters are the same as in the case of the $NN$ potential in the previous section. The light quark mass corresponds to $m_\pi \simeq 370$ MeV while the strange quark mass is tuned to reproduce the physical K meson mass, $m_K\simeq 550$ MeV.
Qualitative features of the $YN$ potential are similar to those of the $NN$ potential. Weak attraction appears at long and intermediate distances while a strong repulsive core shows up at short distance.
However  the spin dependence of the potential is stronger than in the $NN$ case. In particular, the repulsive core in the $^1S_0$ channel is much stronger than that in the $^3S_1$ channel.
Although these results are still preliminary, they are interesting and encouraging as a first step.

Lattice QCD calculations of potentials between hadrons from wave functions have just begun and the first result of the $NN$ potential surprisingly reproduces all the known features of the $NN$ potential such as
weak attraction at long distance, a little stronger attraction at intermediate distance and a strong repulsive core at short distance. For a  quantitative comparison between lattice results and experimental ones, however, chiral and continuum extrapolations are necessary to remove systematic errors.
The inclusion of dynamical quark effects in the $NN$ potential will be the most exciting improvement in future calculations. Now  a door is open for us to vast fields in nuclear physics with lattice QCD.

\section*{Acknowledgement}
I would like to thank T. Hatsuda, N. Ishii, N. Ishizuka, H. Isozaki, H. Nemura, K. Sasaki for valuable discussions and for supplying me with data and plots used in this review.
This work is supported in part by the Grant-in-Aid for Scientific Research from the Ministry of Education, Culture, Sports, Science and Technology (Nos. 13135204, 15540251,18340075. ).

\end{document}